\documentclass[aps,prb,twocolumn,showpacs,superscriptaddress]{revtex4-1}

\usepackage{graphicx} 

\newcommand{\mbr}{\mathbf{r}}
\newcommand{\mbk}{\mathbf{k}}
\newcommand{\mbB}{\mathbf{B}}
\newcommand{\mbG}{\mathbf{G}}
\newcommand{\mbR}{\mathbf{R}}

\begin{document}

\title
{Density-operator theory of orbital magnetic susceptibility in periodic insulators}
\author{X.~Gonze}
\email{xavier.gonze@uclouvain.be}
\affiliation{European Theoretical Spectroscopy Facility (ETSF), IMCN/NAPS
Universit\'{e} Catholique de Louvain, B-1348, Louvain-la-Neuve Belgium}
\author{J.~W.~Zwanziger}
\email{jzwanzig@dal.ca}
\affiliation{Departments of Chemistry and of Physics and 
Atmospheric Sciences, and Institute for Research in Materials,
Dalhousie University, Halifax, NS B3H 4J3 Canada}

\date{\today}

\begin{abstract}

The theoretical treatment of homogeneous static magnetic fields in periodic
systems is challenging, as the corresponding vector potential breaks the
translational invariance of the Hamiltonian.  
Based on density operators and perturbation theory, we propose, for insulators,
a periodic framework for the treatment of magnetic fields up to arbitrary
order of perturbation, similar
to widely used schemes for electric fields.
The second-order term delivers a new, remarkably simple, formulation
of the macroscopic orbital magnetic susceptibility for periodic insulators.
We validate the latter expression using a tight-binding model, analytically
from the present theory and numerically from the large-size limit of a
finite cluster, with excellent numerical agreement.

\end{abstract}

\maketitle

\section{Introduction}

The ability to compute the response of periodic systems to homogeneous
electric fields, strain and atomic displacements is a key ingredient in our current
understanding of dielectric materials (also ferroelectrics, piezoelectrics):
from first principles,\cite{Martin04,Baroni01,Gonze97} one obtains easily
the polarization, dielectric constants, piezoelectric
coefficients, phonon band structure, etc.
Compared to the treatment of such responses for molecules, the handling
of periodic boundary conditions has raised numerous challenges:
for instance, the linear potential
associated with a static homogeneous electric field breaks the translational
symmetry.  While such problems have been successfully
addressed for the above-mentioned perturbations, the
treatment of {\em homogeneous magnetic fields} is not as mature. With the
current interest in multiferro\"ic materials,\cite{Spaldin10}
and the long-term interest in magnetic field-based spectroscopies, a unified
framework for all these responses is highly desirable.

The first strategy followed to treat a periodicity breaking was to
consider perturbations with specific commensurate wavevectors and 
corresponding supercells. For atomic displacements, this approach is known
as the ``frozen-phonon'' method.\cite{Martin04}  Homogeneous
magnetic fields can be treated in this spirit, with an artificial modulation,\cite{Cai04}
although (1) working with supercells is CPU time-consuming, (2) the
study of couplings is tedious in this approach.
More powerful formalisms, based on perturbation theory, that do not rely on supercells or 
long-wavelength limits, have been developed for atomic displacements and
electric fields.\cite{Baroni01, Gonze97,KingSmith93, Nunes02,Veithen05}
For phonons, the long-wavelength phases can be factorized, such that a purely
periodic treatment is recovered. For the electric field,
the position operator can be replaced by the differentiation with respect to the wavevector.
The Berry phase approach to the electrical polarization
is probably the most striking consequence of the latter link.\cite{KingSmith93}
Owing to these advances, linear and nonlinear responses can be
addressed, as well as couplings between different perturbations, in 
a purely periodic primitive cell framework.

For homogeneous magnetic fields in insulators the difficulties outlined
above are more severe. We focus only on the orbital coupling; coupling
to spin is not affected by periodicity issues. The presence of the
magnetic field not only breaks the periodicity of the Hamiltonian but also
induces a vector coupling to the electron dynamics.  In a pioneering
work, Mauri and Louie proposed a method to compute the
orbital magnetic susceptibility (OMS) from the long wavelength limit of an
oscillating perturbation.\cite{Mauri96a} The theory has been adapted to the computation
of the chemical shielding tensor and related 
quantities.\cite{Mauri96b,d'Avezac07,Pickard02}
Although relying on perturbation theory concepts to avoid the use of
supercells, this formalism introduces auxiliary oscillating quantities
that are not consistent with the periodicity of the lattice and that break the
rotational invariance of the global system. Thus for example 
the tensorial structure of the expression of the OMS
could not be recovered.\cite{Mauri96a}
In a related alternative approach, by Sebastiani and Parrinello,\cite{Sebastiani01}
the position operator is replaced by localized sawtooth potentials, one for each orbital.
In practice, the use of supercells, needed to deal with the spatially 
localized Wannier functions, is still necessary in this approach.

Recently, a theory of orbital magnetization has been proposed, in 
which only periodic Bloch wavefunctions and Hamiltonian are used,\cite{TVCR05} 
bringing the understanding of 
this (bulk) quantity to the same formal level as the electric polarization.
Based on this result, the orbital magnetoelectric coupling has been derived from
density-matrix perturbation theory.\cite{Essin10} 
No use of a supercell or long-wavelength limit is needed
in this approach.

In the present contribution, we show how to apply the
density operator approach of Ref.~\onlinecite{Essin10}, to  
{\it arbitrary orders of perturbation} in the magnetic field. 
Focusing on the second order in this expansion, we obtain
a new formula for the OMS, based on the first-order
response of the density operator to magnetic field and wavevector, in
a {\it purely periodic} framework. 
The present approach
is compatible with the similar treatment of electric fields, of atomic
displacements and of their couplings, to arbitrary orders.
We then check the theory by considering a
two-dimensional (2D) periodic tight-binding (TB) model, for which, thanks to the
new approach, the OMS can be obtained as an integral
over the Brillouin Zone (BZ) of an analytical expression.  Alternatively, we
numerically solve this model for clusters of increasing size, considering
explicitly the magnetic field. Essentially exact agreement is obtained
between the two approaches.

Alternative OMS formulae were proposed already fifty years
ago,\cite{Blount62} in the context of effective Hamiltonians or TB models.
However, none of them seem compatible with the currently used formalisms
for electric or atomic displacement responses. 
The present approach is actually considerably simpler. Furthermore,
to our knowledge, no OMS
formula for a periodic insulator has ever been validated by comparison
with numerical results on a solvable cluster model in the large size limit.

\section{Theory}

Let us recall the first steps in the approach of Ref.~\onlinecite{Essin10}.
In atomic units, the Hamiltonian $H$ for an 
electron in a vector potential $\mathbf{A(r)}$ is
$H=\frac{1}{2}\left(-i\mathbf{\nabla} - \frac{1}{c}\mathbf{A}\right)^2 +V(\mbr)$,
where $c$ is the speed of light, and
the potential $V$ is periodic for lattice vectors $\mbR$, that is,
$V(\mbr+\mbR)=V(\mbr)$.
Choosing the gauge $\mathbf{A}=\frac{1}{2}\mbB\times \mbr$,
the Hamiltonian does not possess the translation symmetry of the lattice,
but does have magnetic translation symmetry.\cite{Zak64} The kernel of operators ${\cal O}$
possessing this symmetry can be related to a periodic kernel $\bar{\cal O}$:
\begin{equation}
{\cal O}(\mbr_1,\mbr_2)= 
\bar{\cal O}(\mbr_1,\mbr_2)e^{-i\mbB\cdot \mbr_1\times \mbr_2/2c},
\label{Eq:transform}
\end{equation}
with 
$\bar{\cal O}(\mbr_1+\mbR,\mbr_2+\mbR)=
\bar{\cal O}(\mbr_1,\mbr_2)$. Crucially, the periodic 
counterpart of the Hamiltonian
in this approach has no vector potential dependence
(nor magnetic field dependence): 
$\bar{H}=-\frac{1}{2}\mathbf{\nabla}^2 +V(\mathbf{r})$.

The density operator corresponding to an Hamiltonian, 
for an insulator at $0$ Kelvin, can be obtained by
the minimization of the expectation value of the energy 
of the system on the ensemble of idempotent density matrices 
({\it i.e.} $\rho = \rho \rho$) with fixed electron number:
\begin{equation}
E(\mbB)=
\min_{\rho \, \mathrm{idempotent}} \{ \mathrm{Tr}[\rho H(\mbB) ] \} .
 \label{Eq:E(B)}
\end{equation}
At the minimum, $[\rho,H]=0$.
The density operator also possesses magnetic translation symmetry, 
so that Eq.~(\ref{Eq:E(B)}) can be recast in terms of 
$\bar{H}$ and $\bar{\rho}$ provided that the product of the two operators is 
transformed according to Eq.~(\ref{Eq:transform}).
In real space the product of two operators 
${\cal T} = {\cal VW}$ becomes\cite{Zak64,Essin10}
\begin{equation}
\bar{\cal T}(\mbr_1,\mbr_3)
=\int d\mbr_2\bar{\cal V}(\mbr_1,\mbr_2) \bar{\cal W}(\mbr_2,\mbr_3)e^{-i\phi_{123}/\phi_0},
\label{Eq:productbar}
\end{equation}
where 
\begin{equation}
\phi_{123}/\phi_0
=
{\bf B}.
({\bf r}_1\times {\bf r}_2 + 
 {\bf r}_2\times {\bf r}_3 + 
 {\bf r}_3\times {\bf r}_1 
 )/2
  \label{Eq:phi123}
\end{equation}
is proportional to the magnetic flux through triangle 123.
Based on Eq.~(\ref{Eq:productbar}), 
$\bar{\rho}$ is no longer idempotent, 
due to the appearance of the $e^{-i\phi_{123}/\phi_0}$ phase factor.
In Ref.~\onlinecite{Essin10}, $\bar{\rho}$ is expanded to first order in $\mbB$,
before considering the decomposition of the operators in the BZ, 
in order to obtain the orbital magnetoelectric coupling.

We find that the idempotency problem can be avoided by transforming Eq.~(\ref{Eq:productbar}) 
to a combined BZ and primitive cell integral, {\it before} any expansion in $\mbB$.
To obtain this result, we decompose periodic operators 
$\bar{\cal O}(\mbr_1,\mbr_2)$ into operators
that are separately periodic in each argument and characterized by a wavevector:
\begin{equation}
 \bar{\cal O}(\mbr_1,\mbr_2)= \int_{\mathrm{BZ}}  
\frac {d\mathbf{k}}{(2\pi)^3} 
e^{i \mbk\cdot\mbr_1}
\tilde{\cal O}_\mbk(\mbr_1,\mbr_2)
e^{-i \mbk\cdot\mbr_2},
\end{equation}
with $ \tilde{\cal O}_\mbk(\mbr_1+\mbR,\mbr_2)
= \tilde{\cal O}_\mbk(\mbr_1,\mbr_2+\mbR)
=\tilde{\cal O}_\mbk(\mbr_1,\mbr_2)$,
and
$\tilde{\cal O}_\mbk(\mbr_1,\mbr_2) = \tilde{\cal O}_{\mbk+\mbG}(\mbr_1,\mbr_2)$ 
for all reciprocal-lattice vectors $\mbG$.
Under this decomposition, Eq.~(\ref{Eq:productbar}) becomes
\begin{eqnarray}
\bar{\cal T}(\mbr_1,\mbr_3)=
 \int_{\mathrm{BZ}} \frac {d\mbk}{(2\pi)^3}\int_{\Omega_0} d\mbr_2
e^{i \mbk\cdot(\mbr_1-\mbr_3)} 
\nonumber
\\
\times \tilde{\cal V}_\mbk(\mbr_1,\mbr_2)
\tilde{\cal W}_{\mbk+\Delta\mbk(\mbB)}(\mbr_2,\mbr_3),
\label{Eq:product2}
\end{eqnarray}
where $\Omega_0$ is the primitive cell volume and 
$\Delta\mbk=(\mbr_3-\mbr_1)\times\mbB/2c$ is a position- 
and magnetic field- dependent increment of the wavevector 
$\mbk$. Such an increment defines an $\mbr_1$-dependent kernel, and hence, 
an $\mbr_1$-dependent operator. Although unusual, these expressions 
are mathematically well-defined. 
The transformation proceeds with the Taylor expansion of $\tilde{\cal W}$ 
with respect to $\mbB$, followed by multiple integrations by parts 
with respect to $\mbk$. 
This procedure yields the full expansion of 
$\tilde{\cal T}_\mbk$ to all orders in $\mbB$. 
For simplicity, we denote the derivatives with respect to $\mbk$ in direction $\alpha$
by $\partial_\alpha$, omit real-space arguments, and use the summation convention with 
$\varepsilon$ the totally antisymmetric unit tensor. The expansion is then
\begin{eqnarray}
\tilde{\cal T}_\mbk
=\tilde{\cal V}_\mbk\tilde{\cal W}_\mbk 
+\sum_{m=1}^{\infty} \frac{1}{m!} \left(\frac{i}{2c}\right)^m   
\left(\prod_{n=1}^m \varepsilon_{\alpha_n \beta_n \gamma_n } B_{\alpha_n}\right) 
\nonumber
\\
\times (\partial_{\beta_1} \cdots \partial_{\beta_m} \tilde{\cal V}_\mbk ) 
( \partial_{\gamma_1}  \cdots \partial_{\gamma_m} \tilde{\cal W}_\mbk),
\label{Eq:product3}
\end{eqnarray}
and explicitly to second order,
\begin{eqnarray}
\tilde{\cal T}_\mbk&=&\tilde{\cal V}_\mbk\tilde{\cal W}_\mbk
+ (i/2c) (\varepsilon_{\alpha\beta\gamma}B_\alpha)
    (\partial_\beta \tilde{\cal V}_\mbk)( \partial_\gamma \tilde{\cal W}_\mbk) 
\nonumber
\\
&-&\frac{1}{8c^2}
   \left(\prod_{n=1}^2 \varepsilon_{\alpha_n \beta_n \gamma_n } B_{\alpha_n}\right)
    (\partial_{\beta_1} \partial_{\beta_2} \tilde{\cal V}_\mbk ) 
   ( \partial_{\gamma_1}  \partial_{\gamma_2} \tilde{\cal W}_\mbk) 
\nonumber
\\
&+& O(\mbB^3)
 \label{Eq:product4}
\end{eqnarray}
The density-matrix perturbation theory valid for arbitrary orders of perturbation, 
as developed in ${\bf k}$-space for homogeneous electric fields,\cite{McWeeny62}
can now be generalized to magnetic fields.  Defining perturbation orders
by $\lambda$ such that $\mathbf{B}(\lambda)=\lambda\mathbf{B}_{\lambda=1}$, 
we expand a generic quantity $X$ as
$ X({\bf B})=X^{(0)}+\lambda X^{(1)}+\lambda^2 X^{(2)}+ \cdots$.
The periodic counterpart of the Hamiltonian, $\tilde{H}_\mbk$ has no 
magnetic field dependence, as outlined before, hence, $\tilde{H}_\mbk=\tilde{H}_\mbk^{(0)}$
while for $m \neq 0$, $\tilde{H}_\mbk^{(n)}=0$.
The density operator at any order can be decomposed into different blocks, 
acting inside the occupied subspace (denoted $V$ for Valence), or 
inside the unoccupied subspace ($C$
for Conduction), 
or coupling different subspaces $CV/VC$ (we refer to the subspaces obtained at zero $\mbB$). 
As $\tilde{H}_{\bf k}=\tilde{H}_\mbk^{(0)}$ is block-diagonal in these subspaces,
the expansion of the energy yields
\begin{equation}
E^{(n)}= \int_{\rm BZ} \frac {d\mbk}{(2\pi)^3} 
\mathrm{Tr}[ (\tilde{\rho}^{(n)}_{\mbk VV}+\tilde{\rho}^{(n)}_{\mbk CC}) \tilde{H}_{\bf k} ].
\label{Eq:ECCVV}
\end{equation}
The first order term is related to the magnetization $\mathbf{M}$, $E^{(1)}=\mbB\cdot\mathbf{M}$, 
while the second order term is related to the OMS $\chi$,
$E^{(2)}=2\sum_{ij}B_i\chi_{ij}B_j$.

The $\tilde{\rho}^{(n)}_\mbk$ can be found recursively from the $\tilde{\rho}^{(i)}_\mbk$, with $i<n$, 
by an adaptation of the density-matrix perturbation theory of Ref.~\onlinecite{McWeeny62}. 
The block-diagonal parts of density matrices are given by
\begin{eqnarray}
\tilde{\rho}^{(n)}_{\mbk VV}&=&-\tilde{\rho}^{(0)}_\mbk\tilde{\rho}^{(n)}_{\mbk D} \tilde{\rho}^{(0)}_\mbk,
\label{Eq:rhonVV} \\
\tilde{\rho}^{(n)}_{\mbk CC}&=&
(1- \tilde{\rho}^{(0)}_\mbk)\tilde{\rho}^{(n)}_{\mbk D} (1-\tilde{\rho}^{(0)}_\mbk) ,
\label{Eq:rhonCC}
\end{eqnarray}
where, for first and second orders,
\begin{eqnarray}
\tilde{\rho}^{(1)}_{\mbk D}&&= \frac{i}{2c}\varepsilon_{\alpha\beta\gamma}B_\alpha  
(\partial_{\beta} \tilde{\rho}^{(0)}_\mbk ) 
(\partial_{\gamma} \tilde{\rho}^{(0)}_\mbk ),
\label{Eq:rho1D}
\\
\tilde{\rho}^{(2)}_{\mbk D}&&=\tilde{\rho}^{(1)}_\mbk\tilde{\rho}^{(1)}_\mbk 
\nonumber
\\
+
\frac{i}{2c} 
&&
\varepsilon_{\alpha\beta\gamma}B_\alpha
   [ ( \partial_\beta \tilde{\rho}^{(0)}_\mbk) (\partial_\gamma \tilde{\rho}^{(1)}_\mbk) 
   +
      ( \partial_\beta \tilde{\rho}^{(1)}_\mbk)   (\partial_\gamma \tilde{\rho}^{(0)}_\mbk)
   ] 
\nonumber
\\
-
\frac{1}{8c^2}
&&
   \left(\prod_{n=1}^2 \varepsilon_{\alpha_n \beta_n \gamma_n } B_{\alpha_n}\right)
    \partial_{\beta_1} \partial_{\beta_2} \tilde{\rho}^{(0)}_\mbk .
    \partial_{\gamma_1}  \partial_{\gamma_2} \tilde{\rho}^{(0)}_\mbk.
    \label{Eq:rho2D}
\end{eqnarray}
For the full $\tilde{\rho}^{(1)}$, the off-diagonal $CV$ blocks are needed in addition 
to the first-order $CC$ and $VV$ blocks, and are
obtained by solving the equation
\begin{eqnarray}
[\tilde{H}_\mbk ,\tilde{\rho}^{(1)}_{\mbk CV}]=
\frac{i}{2c}\varepsilon_{\alpha\beta\gamma}B_\alpha(1-\tilde{\rho}^{(0)}_\mbk)  
\nonumber
\\
\times [
(\partial_{\beta} \tilde{\rho}^{(0)}_\mbk ) (\partial_{\gamma} \tilde{H}_\mbk)
-
(\partial_{\beta} \tilde{H}_\mbk ) (\partial_{\gamma} \tilde{\rho}^{(0)}_\mbk )
] \tilde{\rho}^{(0)}_\mbk,
\label{Eq:rho1CV}
\end{eqnarray}
where the r.h.s expression is a $CV$ projection
of the angular-momentum operator times the magnetic field, in reciprocal space.\cite{McWeeny62}

For $n = 1$, these equations reduce
to Eqs.~(21), (23), and (28) of Ref.~\onlinecite{Essin10}, 
where they have been further expressed in terms of Bloch wavefunctions for {\it occupied states} and their first 
$k$-derivatives. When substituted into the first-order term of Eq.~(\ref{Eq:ECCVV}),
the multi-band formula derived in Ref.~\onlinecite{TVCR05} for the magnetization is obtained 
(as $\mathbf{B\cdot M}$). 
Likewise, the second-order equations
can be expressed in terms of the Bloch wavefunctions for occupied states only, their first-order changes, 
and their first- and second- $k$-derivatives, 
yielding an explicitly periodic formulation of the OMS. 
These rather lengthy expressions will be detailed elsewhere.
The tensorial structure of Eqs.~(\ref{Eq:ECCVV}--\ref{Eq:rho1CV}) is obvious, as the terms depend
on $n$ factors of $B$ in different directions.

The OMS is seen from Eq.~(\ref{Eq:ECCVV}) to arise from
both the valence and the conduction subspaces, like the magnetization.
For both subspaces, from Eq.~(\ref{Eq:rho2D}), there are three
contributions: a term quadratic in the density-operator response
$\tilde{\rho}^{(1)}_{\bf k}$, a term linear in this response,
and a term that is independent of
the response of the electrons.  The quadratic and linear $CV$ contributions
can be linked to each other on the basis of Eq.~(\ref{Eq:rho1CV}).
Their sum, being always negative and due to density matrix relaxation,
yields the Van Vleck 
paramagnetic contribution in the present 
formalism.\cite{Splitting}
Alternative decompositions of the OMS
Eq.~(\ref{Eq:rho2D}) exist, similarly to the different expressions for
the dielectric susceptibility in density functional perturbation theory,
see, {\em e.g.}, Eqs.~(37) and (38) of Ref.~\onlinecite{Gonze97}.  
The expression for the OMS presented here
can be shown to be variational, and delivers Eq.~(\ref{Eq:rho1CV})
from Euler-Lagrange conditions.
%

\section{Validation of Theory}

\begin{figure}
\caption{\label{tbmodel}Tight-binding model used to check the theory. A 
square lattice of $A$ sites and $B$ sites are coupled by energies $t$ (A-B
nearest-neighbors) and $s$ (A-A nearest neighbors).}
\includegraphics{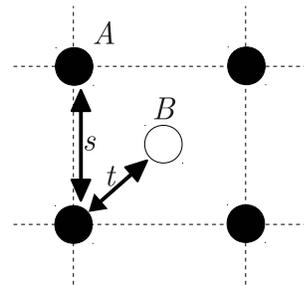}
\end{figure}

The predictions of the theory were checked using a tight-binding
model (Fig.~\ref{tbmodel}).
This model consists of a 2D square lattice with $A$ 
sites at the vertices and $B$ sites at the center of each square, with
on-site energies $E_A < E_B$. Nearest-neighbor $A,B$ pairs are coupled by
energy $t$, and nearest-neighbor $A,A$ pairs by energy $s$ (Fig.~\ref{tbmodel}).
The $A$ sites are defined as initially occupied and the B sites empty. 
The two different types of couplings are necessary to exhibit 
non-trivial behavior from all three contributions
to the susceptibility; with only the more obvious $t$ couplings, only the 
frozen-density-operator contribution is non-zero.
Following Ref.~\onlinecite{Essin10}, the presence of a magnetic field perpendicular to the
plane of the model is included by multiplying the Hamiltonian matrix elements
by $e^{-iB(\mbr_a\times \mbr_b)/2}$,  where 
$\mbr_a$ and $\mbr_b$ describe the locations of the coupled sites.

This model is simple enough for Eqs.~(\ref{Eq:ECCVV}--\ref{Eq:rho1CV}) 
to lead to analytical expressions, apart for a global integral over the 2D BZ. 
The occupied (valence) band eigenenergies $E_{\mbk v}$ are given by
\begin{equation}
E_{\mbk v}=\bar{E}_\mbk-D_\mbk,
\end{equation}
with 
\begin{eqnarray}
D_\mbk&=&\sqrt{(1+\bar{E}_\mbk)^2+\Delta^2_\mbk},\\
\bar{E}_\mbk&=&-s (\cos k_x + \cos k_y),\\
\Delta_\mbk&=&-4t \cos(k_x/2)\cos(k_y/2).
\end{eqnarray}
The periodic part of the Bloch eigenfunctions for the valence band
are
\begin{equation}
|u_{\mbk v}\rangle=\cos \theta_\mbk |A\rangle + \sin \theta_\mbk |B\rangle,
\end{equation}
where
\begin{equation}
\tan \theta_\mbk=(1-\bar{E}_\mbk -D_\mbk)/\Delta_\mbk.
\label{Eq:theta}
\end{equation}
With these definitions, the frozen, linear, and quadratic density operator contributions 
to $\chi$ are respectively
\begin{eqnarray}
\chi_{\mathrm{froz}}  
&=& 
\int_{\rm BZ} \frac {d\mbk}{4(2\pi)^3}  D_\mbk  \left[ 
\left(\frac{\partial^2 \theta_\mbk}{\partial k_x  \partial k_y} \right)^2 
- \frac{\partial^2 \theta_\mbk}{\partial k_x^2} \frac{\partial^2 \theta_\mbk}{\partial k_y^2}\right], \nonumber  \\
\chi_{\mathrm{lin}} &=& -2 \chi_{\mathrm{quad}} 
= -2 (2\pi)^{-3} \int_{\rm BZ} d\mbk R^2_\mbk D_\mbk^{-5},
\end{eqnarray}
with
\begin{eqnarray}
R_\mbk &=&
st(1+\bar{E}_\mbk) \sin(\frac{k_x}{2})\sin(\frac{k_y}{2}) \nonumber \\
&&\times 
[\cos^2(\frac{k_y}{2})-\cos^2(\frac{k_x}{2})] .
\end{eqnarray}
These integrals were carried out numerically to $10^{-10}$ convergence.

The susceptibility of the model was also computed directly, by
diagonalization of the Hamiltonian matrix for finite size lattices.
For an $N\times N$ grid the
energy converged roughly like $1/N$. 
For a given parameter triple $(t,s,B)$ the energy
was computed for a range of lattice sizes ($N=10 \cdots 100$), and the resulting values fit to
a fourth order polynomial in $1/N$. 
This procedure yielded an estimate of the infinite-size limit 
accurate to about 1 part in $10^6$. 
To compute the second order energy change with magnetic field, energies
as a function of mesh size were computed for parameter triples
$(s,t,B)$, where $B$ was one of $0.00$, $0.05$, and $0.10$, and $s$ and
$t$ were fixed.  Then $E^{(2)}$ was estimated
from a finite-difference formula, valid when $E(B)=E(-B)$: 
$E^{(2)} \approx [16 E(B/2)-E(B)-15 E(0)]/(3B^2)$, for 
$B=0.10$.

\begin{figure} 
\caption{\label{t2s}Magnetic susceptibility
for the tight-binding model with (a) $t=2.0$ and a range of $s$ values, and
(b) $s=0.2$ and a range of $t$ values. In both cases,
the full theory (solid line) is composed of three terms (dashed line: frozen wavefunction;
dotted: linear in $\tilde{\rho}^{(1)}$; dash-dot: quadratic in
$\tilde{\rho}^{(1)}$). Squares: exact diagonalization values.}
\includegraphics{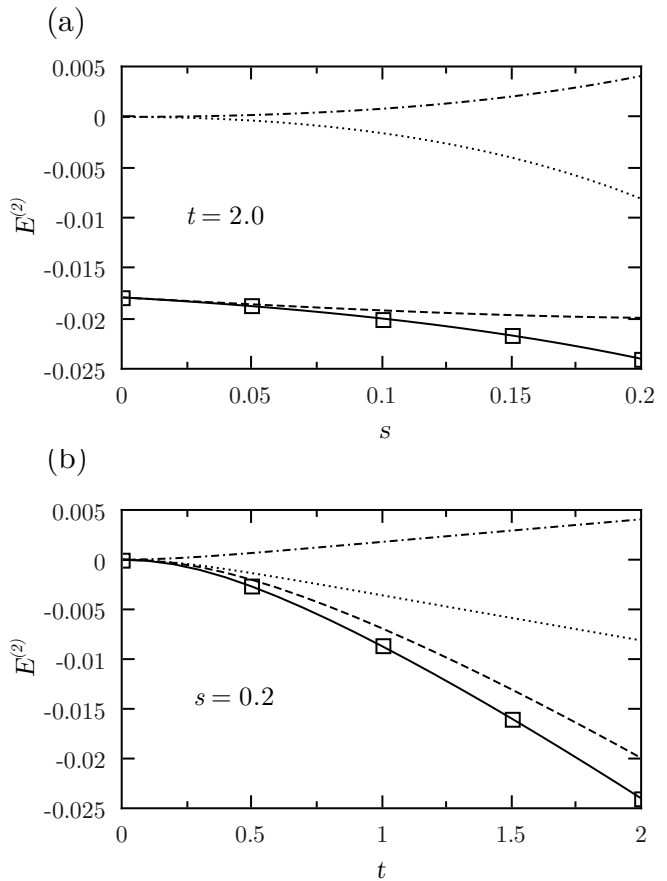}
\end{figure}

Fig.~\ref{t2s} shows the agreement between the theoretical prediction of
$E^{(2)}$ and that obtained by exact diagonalization.
Note that at large $s$ or $t$, 
$E^{(2)}$ has substantial contributions from the linear and quadratic
components, in addition to the frozen term. In all cases agreement between theory
and numerical diagonalization is on the order of a few parts in $10^4$
or better. 

Beyond the OMS, the knowledge of the first-order
density matrix allows computation of {\it every} 
coupling between a magnetic field and other (regular) perturbations
of the system, as well as the orbital current as needed for the nuclear magnetic resonance 
(NMR) shielding. 
Indeed, the mixed derivative of the energy with respect to the magnetic field 
(indexed by $\lambda$) and another perturbation 
(indexed by $\mu$) is given by
\begin{equation}
\frac{\partial^2 E}{\partial \mu \partial \lambda}= 
\int_{\rm BZ} \frac {d\mbk}{(2\pi)^3}
\mathrm{Tr}[ \tilde{\rho}^{(1)}_{\mbk} \frac{\partial \tilde{H}_{\bf k}}{\partial \mu} ].
\label{Eq:mixed}
\end{equation}
where, crucially, the derivative of the first-order density matrix 
with respect to $\mu$ is not needed.

\section{Conclusions}

The present approach to the OMS through expansion
of the total energy, Eq.~(\ref{Eq:ECCVV}),
places it in a new framework, in which its link with the orbital magnetization (from
the first-order
term) is clear, and higher-order susceptibilities can be computed
as well. Of course, the interest in higher-order derivatives of the total
energy with respect to the magnetic field is purely academic. However,
electric fields and vibrational effects can be treated on the same footing, 
at linear and non-linear orders, so that
the path is opened to a unified approach to coupled magnetic, electric and
thermodynamic effects in insulators, expected to help understanding
multiferro\"ics, materials for spintronics applications, as well
as temperature-dependent responses to magnetic fields, as needed for
NMR experiments. 

Although we have focused on insulators at zero Kelvin, 
a generalization to metals at finite temperatures likely exists, as
in the case of the orbital magnetization (the $n=1$ term in our expansion), as outlined in
Ref.~\onlinecite{Shi07}. Note, however, that the density matrix idempotency
relationship $\rho = \rho \rho$ is not valid for such cases, so the density-matrix
perturbation theory as developed in Ref.~\onlinecite{McWeeny62} cannot be applied
straightforwardly.

\section*{Acknowledgments}

X.~G. acknowledges financial support from 
the Walloon Region (WALL-ETSF), the Communaut\'e Fran\c{c}aise de Belgique
(ARC 07/12-003), the Belgian State - IAP Program (P6/42).
J.~Z. acknowledges financial support from the Canada Research Chairs program.

\end{document}